\documentclass[11pt,a4paper]{article}
\usepackage[whole]{bxcjkjatype}

\usepackage{amsfonts,amsmath,amsthm}
\usepackage{graphicx}
\usepackage[colorlinks=true,filecolor=blue,linkcolor=blue,urlcolor=blue,citecolor=blue,linktocpage=true,unicode]{hyperref}
\usepackage{CJKutf8}
\usepackage{tikz}
\usetikzlibrary{knots}
\usetikzlibrary{shadows.blur}
\usetikzlibrary{shadows,arrows}
\usetikzlibrary{shapes,snakes}

\newcommand{\Ysf}{\mathsf{Y}}
\newcommand{\Tsf}{\mathsf{T}}

\newcommand{\eqrefm}[1]{ (\ref{#1}) }
\newcommand{\NS}{Nekrasov--Shatashvili }
\newcommand{\SW}{Seiberg--Witten }

\newcommand{\arxivlink}[2]{\href{https://arxiv.org/abs/#1}{\tt arXiv:#1 [#2]}}

\newtheorem*{theorem*}{}

\date{\today}

\usepackage[framemethod=TikZ]{mdframed}

\newenvironment{itembox}[1]{\begin{mdframed}[
  roundcorner=5pt,
  skipabove=\topskip,
  frametitleaboveskip=\dimexpr-0.7\baselineskip,
  innertopmargin=\dimexpr-0.25\baselineskip,
  innerbottommargin=\dimexpr0.65\baselineskip,
  frametitle={\tikz{\node[anchor=base,rectangle,fill=white]{\strut #1};}}]
  }
  {
   \end{mdframed}}

\begin{document}

\thispagestyle{empty}
\renewcommand{\thefootnote}{\fnsymbol{footnote}}
\setcounter{page}{0}

\begin{titlepage}

 \addtolength{\topmargin}{5em} 
\begin{center}

 \begin{tikzpicture}

  \node at (0,0) {{\LARGE\textbf{Quantum Field Theory:}}};
  \node at (-4,-1.2) {{\LARGE\textbf{Quantum Geometry}}};
  \node at (4,-1.2) {{\LARGE\textbf{Quantum Algebras}}};
  \node at (0,-1.2) {{\LARGE\textbf{and}}};

  \node at (0,-3.5) {{\Large\textbf{---場の量子論：幾何の量子化と量子代数---}}};
 
  \begin{scope}[shift={(0,-6)}]
   \node (D) at (0,0) {\large $\diamondsuit$};
   \node at (-2.1,0) {{\large{\sc Taro Kimura}}};
   \node at (1.5,0) {{\large{\sc 木村太郎}}};  
  \end{scope}

  \begin{scope}[shift={(0,-8)}]

   \node at (-.2,0) {{\large {\it Keio University} \quad $\clubsuit$ \quad 慶應義塾大学}};

   \node at (0,-1.3) 
   {{\large {\it Institute for Basic Science} $\heartsuit$
   \begin{CJK}{UTF8}{nanummj}기초과학연구원\end{CJK}
   $\spadesuit$ 基礎科學硏究院}};   
   
  \end{scope}
  
 \end{tikzpicture}
 \end{center}

 \vskip1cm
\begin{abstract} 
 We demonstrate how one can see quantization of geometry, and quantum algebraic structure in supersymmetric gauge theory.
\end{abstract}
\end{titlepage}

\renewcommand{\thefootnote}{\arabic{footnote}}
\setcounter{footnote}{0}

\vfill\eject

\tableofcontents
\hrulefill
\vskip1em

 \section{はじめに}

場の理論は言うまでもなく理論物理学における最も普遍的な方法の１つですが，一方で必然に無限の自由度を伴うため数学的に満足のいく形での定式化はなされていませんし，ましてそれを一般に解くことは出来ません．
ですが，そうした中でも可解な理論はいくつかあり，低次元の理論や高い対称性を持つ理論などではその真空 (基底状態) やスペクトルを厳密に決定することが出来る場合があります．
拡大された超対称性を持つゲージ場の理論はそうした可解な理論の１つであり，例えば 4 次元 $\mathcal{N}=2$ 超対称ゲージ理論の真空は \textbf{\SW 理論} と呼ばれる幾何学的な手法によって厳密に決定されます．
近年，こうした幾何的対象の量子化と，それに付随する量子的な代数構造が数理物理分野の様々な文脈で登場しており，１つの中心的話題になっています．
本稿では特に超対称ゲージ理論を例として，場の理論から幾何の量子化と量子代数が現れる様子を紹介したいと思います.%

\section{Seiberg--Witten 理論：古典論}

まず初めに \SW 理論について簡単におさらいしておきます~\cite{I}．
代数曲線とは 2 変数有理関数 $H(x,y)$ の零点として与えられるものです：
\begin{align}
 \Sigma = \{\, (x,y) \mid H(x,y) = 0 \,\}
\end{align}
例えば $(x,y)$ を実数として $H(x,y) = x^2 + y^2 - 1$ とすれば原点を中心とする単位円で 1 次元ですが，複素数で考えると代数曲線は複素 1 次元，つまり実 2 次元になります．
\SW 理論によれば 4 次元 $\mathcal{N}=2$ SU($N$) ゲージ理論の場合には \textbf{\SW 曲線}と呼ばれる
\begin{align}
 H(x,y) = y + \frac{1}{y} - T_N(x)
 \label{eq:H-func_A1}
\end{align}
で与えられる代数曲線で真空の Coulomb 枝を決定することが出来ます．
ただし $T_N(x) = x^N + \cdots$ は $N$ 次多項式で，代数曲線としては Fig.~\ref{fig:curve} の様に種数 $N-1$ の超楕円曲線になります．

\begin{figure}[t]
 \begin{center}
  \begin{tikzpicture}[thick,scale = 2]

   \draw 
   (-.5,0) to [out=down,in=left]
   (1,-.6) to [out=right,in=left]   
   (2,-.5) to [out=right,in=left]
   (3,-.6) to [out=right,in=left]
   (4,-.5) to [out=right,in=left]
   (5,-.6) to [out=right,in=down]
   (6.5,0)   to [out=up,in=right]
   (5,.6)  to [out=left,in=right]
   (4,.5)  to [out=left,in=right]
   (3,.6)  to [out=left,in=right]
   (2,.5)  to [out=left,in=right]
   (1,.6)  to [out=left,in=up]  (-.5,0);

   \foreach \x in {.7,3,5.3}{
       
    \begin{scope}[shift={(\x,-.17)}]

     \draw [clip] (-.5,.3) to [bend right] (.5,.3);
     \draw (-.5,0.1) to [bend left] (.5,0.1);
    
    \end{scope}

   \begin{scope}[shift={(\x,-.17)}]

     \draw[red,rotate = 70] (.12, .1) arc [start angle = 0, end angle = 180, x radius = .3, y radius = .1];

   \draw[red,-latex, rotate = 70] (.12, .1) arc [start angle = 0, end angle = 120, x radius = .3, y radius = .1];

   \draw[red,dotted, rotate = 70] (.12, .1) arc [start angle = 0, end angle = -180, x radius = .3, y radius = .1];       

    \draw[blue] (.8,.3) arc [start angle = 0, end angle = 360, x radius = .8, y radius = .3];
    \draw[blue,-latex] (.8,.3) arc [start angle = 0, end angle = 320, x radius = .8, y radius = .3];
    
    \end{scope}

   }
   
   \node at (.5,-.9) {$A_1$};
   \node at (2.7,-.9) {$A_2$};
   \node at (5,-.9) {$A_3$};   

   \node at (1.5,-.3) {$B_1$};
   \node at (3.8,-.3) {$B_2$};
   \node at (6.,-.3) {$B_3$};   
   
  \end{tikzpicture}
 \end{center}
 \caption{SU(4) ゲージ理論を記述する種数 3 の超楕円曲線．合計 6 つの非自明な周期がある．}
 \label{fig:curve}
\end{figure}

次に，こうした幾何的対象から物理量を取り出すために以下の様な 1 次微分形式 (\SW 微分) を曲線 $\Sigma$ 上に定めます：
\begin{align}
 \lambda = x \frac{dy}{y} = x \, d (\log y)
 \label{eq:SW_diff}
\end{align}
すると種数 $N-1$ の曲線には $2N-2$ 個の非自明な周期 $(A_i,B_i)_{i=1,\ldots,N-1}$ があるので，それぞれ 1 次形式の周期積分として
\begin{align}
 \oint_{A_i} \lambda = a_i
 \,, \quad
 \oint_{B_i} \lambda 
 = \frac{\partial \mathcal{F}}{\partial a_i}
 \label{eq:period_int}
\end{align}
が得られます．
ここで $a_i$ は Coulomb モジュライと呼ばれる真空を特徴付ける座標で，$\mathcal{F}$ はプレポテンシャルと呼ばれる正則関数です．
さらに低エネルギーでの有効的な結合定数は周期行列 $\tau_{ij} = \partial_i \partial_j \mathcal{F}$
で決まります．
非可換ゲージ理論は漸近的自由性から低エネルギーでは強結合になるので一般に量子補正を決めることは難しいのですが，\SW 理論ではこの様な幾何的な性質で決まってしまう，というのが著しい特徴です．

\section{Seiberg--Witten 理論の量子化：可積分系とのつながり}

\SW 理論は当初より発見法的で，うまくいくことは分かっていても，なぜそうなるのか，という点で疑問がありましたが，その後 Nekrasov による \SW 理論の直接導出が１つの契機になりました．
場の理論の経路積分は一般に無限次元の複雑な積分であり，例えば相互作用の弱い場合には摂動的な解析が用いられますが，相互作用が強くなると摂動論も破綻してしまいます．
ところが超対称性がある場合などには Duistermaat--Heckman の局所化と呼ばれる方法が適用可能で，その場合には無限次元の経路積分が有限の積分や離散和にまで簡単化します．
$\mathcal{N}=2$ 超対称ゲージ理論も経路積分で定められる分配関数が局所化によって離散和として表されますが，その組み合わせ論的な表式は Nekrasov 分配関数と呼ばれており，対称群の既約表現に対する Plancherel 測度の一般化になっています．

Euclid 4 次元空間は $\mathbb{R}^4 = \mathbb{C}_1 \times \mathbb{C}_2$ より複素 2 次元空間ですが，それぞれの複素平面の回転に対応するパラメータ $(\epsilon_1,\epsilon_2)$ を導入することが可能で，
$\Omega$ 背景場 (あるいは重力光子場) と呼ばれます．
Nekrasov 分配関数には $(\epsilon_1,\epsilon_2)$ の依存性があり，その漸近的な振る舞いから \SW 理論のプレポテンシャルが導かれます：
\begin{align}
 \epsilon_1 \epsilon_2
 \log \mathcal{Z}(\epsilon_1,\epsilon_2)
 =: \mathcal{F}(\epsilon_1,\epsilon_2)
 \stackrel{\epsilon_{1,2}\to 0}{\longrightarrow} \mathcal{F}
\end{align}
つまり，これによって \SW 理論が場の理論の経路積分から導かれたことになります．
以下では $\epsilon_{1,2} \to 0$ の極限を \SW 極限と呼ぶことにします．
この $\Omega$ 背景は当初，分配関数を正則化するために導入されたものでしたが，その後，発想を逆転してむしろ $\Omega$ 背景を理論の変形，つまり量子化に用いようという試みが提案されました．
その１つが \NS のアイデアで \SW 理論を可積分系として捉えようというものです．

\SW 理論の提案の直後から，\SW 曲線が戸田格子や楕円 Calogero--Moser 系などの古典可積分系のスペクトル曲線と同定出来ることが指摘されていましたが，この対応を量子系へ持ち上げることは出来るのでしょうか．
量子可積分系の場合には Baxter の TQ 関係式や Bethe 仮説方程式などがスペクトル曲線の量子化に対応します．
\NS は $\Omega$ 背景を \SW 極限 $\epsilon_{1,2} \to 0$ でなく，$\epsilon_1$ を有限に固定したまま $\epsilon_2 \to 0$ の極限を取ると，その場合のゲージ理論の真空条件が Bethe 仮説方程式で与えられることを示しました~\cite{NS}．
この時，残った $\epsilon_1$ が量子可積分系の Planck 定数 $\hbar$ に対応しており，正に $\Omega$ 背景によって \SW 理論が量子化されたことになります．
以下でもう少し詳細に量子化の様子を見てみましょう．

\NS 極限では前述のプレポテンシャルでなく，以下に定められる超ポテンシャルを考えます：%
\begin{align}
 \epsilon_2
 \log \mathcal{Z}(\epsilon_1,\epsilon_2)
 =: \mathcal{W}(\epsilon_1,\epsilon_2)
 \stackrel{\epsilon_{2}\to 0}{\longrightarrow} \mathcal{W}(\epsilon_1)
\end{align}
ここでゲージ理論の真空条件 ($F$ 項条件) は
\begin{align}
 \exp \left(\frac{\partial \mathcal{W}}{\partial a_i} \right)
 = 1
\end{align}
と与えられ，つまり超ポテンシャルの微分が $(2 \pi i)^{-1} \partial \mathcal{W}/\partial a_i \in \mathbb{Z}$ として整数値に量子化されます．
すると周期積分\eqrefm{eq:period_int}は
\begin{align}
 \frac{1}{2 \pi i}
 \oint_{B_i} \lambda
 &
 = \frac{\epsilon_1}{2 \pi i} \, \frac{\partial \mathcal{W}}{\partial a_i}
 = \epsilon_1 \times \mathbb{Z}
\end{align}
より $\epsilon_1$ を単位として量子化します．
ここで 
\begin{align}
 \frac{\partial \mathcal{F}}{\partial a_i}\Bigg|_{\epsilon_2 \to 0}
 \hspace{-.5em}
 =
 \lim_{\epsilon_2 \to 0} \epsilon_1 \epsilon_2 \frac{\partial \log \mathcal{Z}}{\partial a_i}
 =
 \epsilon_1 \frac{\partial \mathcal{W}}{\partial a_i}
\end{align}
となることを用いました．
この様な周期積分の量子化は前期量子論の Bohr--Sommerfeld 量子化を彷彿とさせます：
\begin{align}
 \frac{1}{2 \pi i}
 \oint p dx = i \hbar \times \mathbb{Z}
\end{align}
この周期積分はシンプレクティック 2 形式 $\omega = d (pdx) = dp \wedge dx$ を用いて面積分 $\displaystyle \oint p dx = \int \omega$ として書けますが，
これを演算子形式にして局所的にしたものが正準交換関係による量子化条件です：
\begin{align}
 [\hat{x}, \hat{p}] = i \hbar
\end{align}
この手続きをゲージ理論に適用すると \SW 微分 (\ref{eq:SW_diff}) の周期積分が Bohr--Sommerfeld の意味で量子化しているので，シンプレクティック 2 形式
\begin{align}
 \omega = d \lambda = d x \wedge d \log y
\end{align}
より，正準交換関係
\begin{align}
 \left[ \hat{x}, \log \hat{y}\right] = \epsilon_1
\end{align}
を得ます.
ここでは 4 次元を考えていますが，5 次元理論では \SW 微分は $\lambda = \log x \, d (\log y)$ で与えられるので，対応する 2 形式は $\omega = d \log x \wedge d \log y$ です．
従って正準交換関係は
\begin{align}
 [\log \hat{x}, \log \hat{y}] = \epsilon_1
 \label{eq:CCR}
\end{align}
となり $(x,y)$ の役割が対等になります．実際にこの $(x,y)$ の入れ替えは弦理論的には S 双対性として理解できます．
以下では都合上 5 次元 (K 理論) の表記 $(q_1,q_2) = (e^{\epsilon_1},e^{\epsilon_2})$ を用います．
また $q = q_1 q_2$ と略記し，$\Omega$ 背景が自己双対 $\epsilon_1 + \epsilon_2 = 0$ の場合には $q = 1$ に対応します．

ここまで周期積分の量子化から始めて正準量子化の手続きまでしましたが，ゲージ理論としてはどの様にこの正準交換関係が見えるでしょうか．
一度 \SW 曲線にまで立ち戻ってみましょう．
\SW 曲線は有理関数\eqrefm{eq:H-func_A1}の零点で与えられるので
\begin{align}
 y(x) + \frac{1}{y(x)} = T_N(x)
 \label{eq:alg_rel1}
\end{align}
と等価です．
ここで $y(x)$ というのはこの代数関係式を用いて $y$ を $x$ の関数として考える，という意味ですが，ゲージ理論における $\mathsf{Y}$ 演算子と呼ばれる演算子の期待値 $\langle \, \mathsf{Y}(x) \, \rangle = y(x)$ を考えると，\SW 極限において実際に関係式\eqrefm{eq:alg_rel1}を満たすことが示せます．
つまり \SW 曲線は 1 点相関関数に対する代数関係式としても解釈出来ます．
次に \NS 極限の下で 1 点関数 $y(x)$ の満たす関係式を見てみましょう．
具体的に計算すると次の様な関係式を得ます：
\begin{align}
 y(x) + \frac{1}{y(q_1^{-1} x)} = T_N(x; \epsilon_1) 
 \label{eq:alg_rel2} 
\end{align}
右辺は再び変数 $x$ の多項式ですが，その係数には $\epsilon_1$ の依存性があります．
また左辺を見てみると 1 点関数 $y(x)$ の引数が $q_1$ だけシフトしています．
つまり今の場合には相関関数が満たすのは代数関係式でなく，差分方程式になっているということです．
さらに変数 $y(x)$ を
\begin{align}
 y(x) = \frac{Q(q_1 x)}{Q(x)} 
 \label{eq:log-diff}
\end{align}
の様に書いてみましょう.%
\footnote{%
この変数変換は Riccati 型微分方程式を線形化する際の変数変換 $\phi(x) = \psi(x)'/\psi(x)$
の差分版と解釈出来ます．
}
すると元の差分方程式\eqrefm{eq:alg_rel2}から $Q(x)$ に対する差分方程式
\begin{align}
 Q(q_1 x) + Q(q_1^{-1} x) = T_N(x:\epsilon_1) Q(x)
 \label{eq:TQ1}
\end{align}
を得ます．
これは量子 SU(2) XXX スピン鎖模型に対する TQ 関係式 (の退化版) に他なりませんので，Bethe 仮説方程式も $Q(x)$ の零点，すなわち Bethe 根を代入することで直ちに得られます．
さらにこの関係式は
\begin{align}
 \left[
 \hat{y} + \hat{y}^{-1} - T_N(\hat{x};\epsilon_1)
 \right] Q(x) = 0
 \label{eq:TQ2}
\end{align}
の様に Schr\"odinger 型の差分方程式に書き直せます．
ここで $\hat{x} = x$ で $\hat{y}$ は差分演算子です：
\begin{align}
 \hat{y}^{\pm 1} = \exp \left( \mp \epsilon_1 \frac{\partial}{\partial \log x}\right)
\end{align}
するとこれらの $(\hat{x},\hat{y})$ は正準交換関係\eqrefm{eq:CCR}を満たしますので，この意味で本来の代数関係式\eqrefm{eq:H-func_A1}を量子化して得られたのが TQ 関係式ということになります：
\begin{align}
 H(\hat{x},\hat{y}) Q(x) = 0
\end{align}
この様に2変数関数 $H(x,y)$ の零点として与えられていた代数曲線 $\Sigma$ に対して，正準量子化を施して得られた演算子 $H(\hat{x},\hat{y})$ 
として特徴付けられるものを\textbf{量子曲線 (quantum curve)} と呼びます．
つまり $\Omega$ 背景を導入することによって \SW 曲線を量子化することが出来たわけです．

\section{\SW 理論の 2 重量子化：演算子形式}

前節で論じたのはゲージ理論の \NS 極限で，$\Omega$ 背景場によって \SW 曲線が量子化される様子を見ました．
この時 $\Omega$ 背景の内，片方を零にする極限を議論しましたが，元来 2 つのパラメータは対等でした.%
\footnote{$(\epsilon_1,\epsilon_2)$ を入れ替える操作は Langlands 双対を取る操作に対応しています．従ってこれらの等価性は通常のゲージ理論が後述の箙ゲージ理論として simply-laced であることを意味します．また最近は non-simply-laced な箙ゲージ理論の構成も与えられています．}
そうすると 2 つのパラメータのいずれも量子化パラメータとして機能するのでは？とか，2 つを同時に使うことでさらなる量子化が可能なのでは？という疑問が出てきます．
そこで次に，一般の $\Omega$ 背景の下での量子化について見ていきたいと思います．

変数 $y(x)$ が演算子 $\mathsf{Y}(x)$ の期待値で表されることを思い出しましょう．
すると一般の $\Omega$ 背景 $(\epsilon_1, \epsilon_2)$ の下では $\mathsf{Y}$ 演算子の 1 点関数に対して次の様な関係が成り立ちます：
\begin{align}
 \left<
  \mathsf{Y}(x) + \frac{1}{\mathsf{Y}(q^{-1} x)} 
 \right>
 =
 T_N(x;\epsilon_1,\epsilon_2)
 \label{eq:qq-ch_A1}
\end{align}
ここで右辺は再び変数 $x$ の $N$ 次多項式で，その係数には $(\epsilon_1, \epsilon_2)$ の依存性があります．
また \SW 極限や \NS 極限では $\left< \mathsf{Y}(q^{-1}x)^{-1} \right> = y(q^{-1}x)^{-1}$ となりますが，一般の $\Omega$ 背景では成り立たないことに注意して下さい．
これにより \SW 極限・\NS 極限での関係を再現します．

ここで左辺は $\Ysf$ 演算子たちの期待値で与えられていますが，この表式から $\left< \mathsf{T}(x) \right> = T_N(x;\epsilon_1,\epsilon_2)$ となる様な演算子 $\mathsf{T}$ が自然に導入されます．
つまり関係式\eqrefm{eq:qq-ch_A1}を $\Tsf$ 演算子の定義式として見ることが出来ます：
\begin{align}
 \mathsf{T}(x) := \mathsf{Y}(x) + \frac{1}{\mathsf{Y}(q^{-1} x)}
 \label{eq:qq-ch_A1_op} 
\end{align}
この $\Tsf$ 演算子が何者であるかを説明するために演算子形式というものを導入しましょう．
まず $\mathcal{N}=2$ 超対称ゲージ理論を特徴付ける (紫外領域での) プレポテンシャルを次の様に正則な演算子を用いて変形します：
\begin{align}
 \mathcal{F} \ \longrightarrow \
 \mathcal{F} + \sum_{n=1}^\infty t_n \, \mathcal{O}_n
 \label{eq:UV_def}
\end{align}
ここで演算子 $\mathcal{O}_n$ はカイラル環演算子で，4 次元だと
複素スカラー場のゲージ不変な演算子 $\mathcal{O}_n = \operatorname{Tr} \Phi^n$ として与えられます．
するとゲージ理論の分配関数には
\begin{align}
 \exp\left( \sum_{n=1}^\infty t_n \, \mathcal{O}_n \right)
\end{align}
という項が付け加わりますが，これは行列模型で言う所のポテンシャル項に相当します．
この項があると演算子 $\mathcal{O}_n$ は共役な結合定数 $t_n$ の微分として実現されます：
\begin{align}
 \mathcal{O}_n \ \longleftrightarrow \ \frac{\partial}{\partial t_n}
\end{align}
つまり $\mathcal{O}_n$ は微分演算子として表され，交換関係 $[\mathcal{O}_n , t_{n'}] = \delta_{n+n',0}$ を満たします．
本稿ではこの様な取り扱いを\textbf{演算子形式}と呼びます．
これは Fourier 変換の際に $e^{i p x / \hbar}$ という平面波項があると $p = - i \hbar \partial_x$ の様に共役変数の微分として表され，正準交換関係 $[x, p] = i \hbar$ を満たす，というのと事情は同じです．
従って，プレポテンシャルの変形\eqrefm{eq:UV_def}によって演算子形式を導入することは，\NS 極限とは別の意味での量子化を行うことに対応します．

ではこの演算子形式の下で，$\Ysf$ 演算子はどの様に表されるでしょうか．
天下り的ではありますが，ゲージ理論の議論から次の様な表式が与えられます：
\begin{align}
 \Ysf(x)
 & =
 \exp
 \left(
  \sum_{n \neq 0} y_n \, x^{-n}
 \right)
\end{align}
ただし演算子 $y_n$ は交換関係
\begin{align}
 \Big[y_n, y_{n'} \Big]
 =
 - \frac{1}{n} \frac{(1 - q_1^n)(1 - q_2^n)}{1 + q^n}
 \delta_{n+n',0}
\end{align}
を満たします．
また簡単のため，零モードと呼ばれる項も省きました．
この交換関係は $\Omega$ 背景パラメータによって変形されてはいますが，通常の調和振動子に現れる演算子と本質的に同じものです．
ただし \NS 極限では可換な代数に帰着されることに注意します．
この様な振動子による表示を共形場理論や可積分系の文脈では\textbf{自由場表示}と呼びます．

続いて $\Tsf$ 演算子です．
すでに $\Ysf$ 演算子の表式が与えられていて，また $\Tsf$ 演算子は\eqrefm{eq:qq-ch_A1_op}によって定められています．
そこで $\Tsf$ 演算子を
\begin{align}
 \Tsf(x) = \sum_{n \in \mathbb{Z}} T_n \, x^{-n}
\end{align}
の様に展開すると，展開係数 $T_n$ は
\begin{align}
 \Big[ T_n, T_{n'}\Big]
 & =
 - \sum_{k=1}^\infty f_k
 \left( T_{n-k} T_{n'+k} - T_{n+k} T_{n'-k} \right)
 \nonumber \\
 &
 - \frac{(1-q_1)(1-q_2)}{1-q} \left( q^n - q^{-n}\right)
 \delta_{n+n',0}
 \label{eq:q-Vir_alg}
\end{align}
を満たすことが分かります．
また構造関数 $f(x)$ は
\begin{align}
 f(x)
 & = \sum_{k=0}^\infty f_k \, x^k
 \nonumber \\
 & = \exp \left(
 \sum_{n=1}^\infty \frac{x^n}{n} \frac{(1-q_1^n)(1-q_2^n)}{1+q^n}
 \right)
\end{align}
と定めます．
ただしこの代数関係を直接導出するよりは，それと等価な生成演算子 $\Tsf(x)$ に対する関係式
\begin{align}
 &
 f\Big(\frac{x'}{x}\Big) \Tsf(x) \Tsf(x')
 - f\left(\frac{x}{x'}\right) \Tsf(x') \Tsf(x)
 \nonumber \\
 & =
 - \frac{(1-q_1)(1-q_2)}{1-q}
 \left( \delta \left(q \frac{x'}{x} \right) - \delta \left(q^{-1} \frac{x'}{x} \right) \right)
\end{align}
を導く方が簡単です．
ここで与えられた $\Tsf$ 演算子の代数関係\eqrefm{eq:q-Vir_alg}は，右辺に非線形項を含むため，$T_n$ は Lie 代数ではありませんが，実はこの代数は共形場理論の無限の対称性を記述する \textbf{$q$ 変形 Virasoro 代数}に他なりません~\cite{S}．
つまり，ゲージ理論の \SW 理論から始めて，2段階の量子化を経て量子代数である $q$-Virasoro 代数まで辿り着いたことになります．
しかし現段階では何故そうした量子代数的構造がゲージ場の理論を通じて幾何の量子化から現れたのかはっきりしません．
次節ではより一般の状況を考え，その背景について迫りたいと思います．

\section{箙ゲージ理論から箙W代数へ}

Virasoro 代数にはより一般の W 代数と呼ばれる枠組みがあります．
W 代数は Lie 代数 $\mathfrak{g}$ から構成することが出来ますが，最も簡単な場合 $\mathfrak{g} = \mathfrak{su}(2)$ が Virasoro 代数に対応します．
すると，量子化の手続きによって一般の $q$ 変形 W 代数を導く様なゲージ理論は何か，という問いが生じますが，これは \textbf{{箙}(えびら; quiver) ゲージ理論}というものを使うことで構成出来ます．
通常のゲージ理論ではゲージ群は１つですが，それを複数のゲージ群と対応するゲージ場を含む様に拡張したのが箙ゲージ理論で，それを特徴付けるのに頂点とそれらを結ぶ辺から成る箙図 (図~\ref{fig:quiver}) を使います．
中にはこの箙図と Lie 代数の分類に用いられる Dynkin 図との類似性に気づかれる読者もいらっしゃるかと思いますが，実際に箙が単純 Lie 代数に対応する Dynkin 図と同じ場合には A 型箙などと呼び，アフィン型に対応する場合には $\widehat{\mathrm{A}}$ 型箙などと表記します．

\begin{figure}[t]
\begin{center}
 \begin{tikzpicture}[thick,scale=1.2]

  \draw (0,0) -- (2,0);
  
  \filldraw[fill=white] (0,0) circle [radius = .1];
  \filldraw[fill=white] (1,0) circle [radius = .1];
  \filldraw[fill=white] (2,0) circle [radius = .1];  

  \begin{scope}[shift={(4.5,0)}]

   \draw (-1,0) -- (0,0) -- (.8,.5);
   \draw (0,0) -- (.8,-.5);
   
   \filldraw[fill=white] (0,0) circle [radius = .1];
   \filldraw[fill=white] (-1,0) circle [radius = .1];
   \filldraw[fill=white] (.8,.5) circle [radius = .1];
   \filldraw[fill=white] (.8,-.5) circle [radius = .1];   
   
  \end{scope}

  \node at (0,1) {(a)};
  \node at (3.5,1) {(b)};  
  
 \end{tikzpicture}
\end{center}
 \vspace{.5em}
 \caption{頂点 (ゲージ群) とそれらを結ぶ辺 (双基本表現場) から成る箙図．(a) $A_3$型 (b) $D_4$ 型箙．}
 \label{fig:quiver}
\end{figure}

この箙ゲージ理論から始めて幾何の量子化の手続きをするためにはまず出発点となる \SW 曲線が必要ですが，実はこの箙ゲージ理論に対する \SW 曲線には箙の構造に付随した興味深い表現論的な性質があります．
例えば SU$(N)$ ゲージ理論の \SW 曲線は代数関係式\eqrefm{eq:alg_rel1}で与えられる代数曲線でした．
ここで流儀にもよりますが，右辺はゲージ群を変えるとそれに応じて変化する一方で，左辺は変化しません．
Nekrasov--Pestun の指摘はこの左辺の組み合わせが SU(2) の基本表現 ($\textbf{2}$ 表現) の指標である，というものです:
\begin{align}
 \chi_{\textbf{2}}(\mathrm{SU}(2)) = y + y^{-1}
\end{align}
この SU(2) の出処は，と言うと，ゲージ群が１つしかない場合は頂点を１つだけ持つ箙に相当しますが，これはいわゆる $A_1$ 箙で，この箙を Dynkin 図と同定した場合に対応する Lie 群は SU(2) である，というところから来ています．
これを一般の箙ゲージ理論の場合に拡張すると以下の様になります：
\begin{itembox}{箙ゲージ理論の \SW 幾何~\cite{NP}}
 $G_\Gamma$ を上述の箙・Dynkin 対応で与えられる単純 Lie 群とすると，$\Gamma$ 箙ゲージ理論の \SW 曲線は $G_\Gamma$ の基本表現指標で与えられる．
\end{itembox}
\noindent
ここで言う基本表現とは，各頂点に対して与えられるウェイトを最高ウェイトとして考え，そこから Weyl 群の作用によって生成される表現のことを指しています．
つまり箙の頂点の数だけ基本表現があります．
またこの処方箋は $\Gamma$ がいわゆる有限型の箙・Dynkin 図でない場合，例えばアフィン型の場合でも適用可能です．
例えば $A_2$ 箙の場合には \SW 曲線は
\begin{subequations} \label{eq:alg_A2}
\begin{align}
 y_1 + \frac{y_2}{y_1} + \frac{1}{y_2} & = T_1(x) \\
 y_2 + \frac{y_1}{y_2} + \frac{1}{y_1} & = T_2(x) 
\end{align} 
\end{subequations}
という連立の関係式で与えられて，例えば $y_2$ を消去して $y_1 = y$ とすると代数曲線の標準的な形
\begin{align}
 H(x,y) = y^3 - T_1(x) \, y^2 + T_2(x)\, y - 1 = 0
\end{align}
を得ます．
ここで $T_{1,2}(x)$ は変数 $x$ の多項式で，その次数は 2 つあるゲージ群の階数に対応しており，また実際に\eqrefm{eq:alg_A2}式の左辺は SU(3) の $\textbf{3}$ 表現と $\bar{\textbf{3}}$ 表現の指標になっていることが確認出来ます．

続いて箙ゲージ理論の \NS 極限での振る舞いを見てみましょう．
この時 \SW 曲線は量子化されて $A_1$ 型箙の場合には\eqrefm{eq:alg_rel2}式の様な差分方程式になりました．
実はこの場合にも表現論的な解釈が可能で，この場合には通常の指標を量子変形した \textbf{$q$ 指標}によって量子 \SW 曲線が与えられます．
この $q$ 指標は Frenkel--Reshetikhin によって量子アフィン代数の有限次元表現に対して導入されたもので，量子可積分系の TQ 関係式や Bethe 仮説方程式を特徴付けます．
従って箙ゲージ理論に対して以下の様な対応を得ます：
\begin{itembox}{箙ゲージ理論の \NS 対応~\cite{NPS}}
 $\Gamma$ 箙ゲージ理論の極限 $(\epsilon_1,\epsilon_2) \to (\hbar, 0)$ における真空条件は量子 $G_\Gamma$ スピン鎖模型の Bethe 仮説方程式で与えられる． 
\end{itembox}
\noindent
例えば $A_2$ 箙ゲージ理論の \NS 極限は $G_{A_2} =$ SU(3) より，SU(3) スピン鎖模型の Bethe 仮説方程式で記述されます．

この様に箙ゲージ理論の \SW 幾何には表現論的な特徴付けがあることが分かりましたが，それでは一般の $\Omega$ 背景の場合にはどうなるでしょうか．
そもそもこの場合に何かしら良い性質があること自体も自明ではありませんが，むしろゲージ理論の $\Omega$ 背景による変形を使って新しい表現論的な対象を定めようという試みが最近なされています．
その 1 つが Nekrasov の提唱する BPS/CFT 対応で，一般の $\Omega$ 背景下では通常の指標や $q$ 指標に代わって \textbf{$qq$ 指標}と呼ばれる指標の 2 重変形が得られます~\cite{N}．
$A_1$ 箙の場合には $\Ysf$ 演算子に対する関係式\eqrefm{eq:qq-ch_A1}に他なりません．
こうして見ると $q$ 指標と $qq$ 指標はただ単に $q_1$ を $q = q_1 q_2$ に置き換えているだけに見えますが，基本表現でなく高次の対称表現や，あるいはより一般の箙を考えると色々と興味深い性質が見えて来ますが，その表現論的な理解はまだ十分でなく，今後の課題です．

最後に，この $qq$ 指標を使って $q$ 変形 W 代数を構成する話を紹介して本稿を締めくくりましょう．
$\Gamma$ 箙ゲージ理論は箙の頂点の数 (ゲージ群の数) だけの基本表現を持ち，それぞれの表現に対して $qq$ 指標が定まります．
この $qq$ 指標を演算子形式で取り扱うことで $\Tsf$ 演算子が与えられますが，この $\Tsf$ 演算子が $q$ 変形 W($\Gamma$) 代数の生成演算子になっていることを示すことが出来ます．
つまり箙の形に対応して様々な W 代数を構成することが出来て，これを\textbf{箙 W 代数}と呼んでいます：
\begin{itembox}{箙 W 代数~\cite{KP}}
 $\Gamma$ 箙ゲージ理論における \SW 曲線の 2 重量子化は $qq$ 指標で与えられ，箙 W 代数 W($\Gamma$) の生成演算子を定める．
\end{itembox}
これまで議論してきた $\Omega$ 背景とゲージ理論・可積分系の対応を Table~\ref{tab:Omega} にまとめました．
2つの $\Omega$ 背景を同時に考えることで \SW 幾何が 2 重量子化され，対応する量子的代数構造として W 代数が現れる，というのが本稿のメッセージです．
この様なゲージ理論と W 代数の関係として他にも AGT 対応と呼ばれる関係が知られていますが，この場合にはゲージ理論のゲージ群に対応した W 代数が得られ，箙 W 代数とは双対の関係にあります~\cite{TT}．

\begin{table}[t]
 \begin{center}
  \begin{tabular}{ccc}\hline\hline
   $\Omega$ 背景 & SW 幾何 & 可積分系 \\\hline
   $(\epsilon_1,\epsilon_2) = (0,0)$ & 指標 & 古典系 \\
   $(\epsilon_1,\epsilon_2) = (\hbar,0)$ & $q$ 指標 & 量子系 \\
   $(\epsilon_1,\epsilon_2) \neq (0,0)$ & $qq$ 指標 & W 代数 \\ \hline\hline
  \end{tabular}
 \end{center}
 \caption{$\Omega$ 背景とゲージ・可積分系対応の量子化．}
 \label{tab:Omega}
\end{table}

この箙 W 代数の特徴をいくつか挙げておくと，1 つは前述の様に，箙 $\Gamma$ は必ずしも有限次元 Lie 代数の Dynkin 図になる必要はない，ということです．
逆に有限型 Dynkin 図に対応しない箙ゲージ理論を用いて新しい W 代数を構成することも出来ます．
例えば最も簡単なアフィン型箙 $\widehat{A}_0$ を考えて見ましょう．
これはゲージ理論としては 4 次元 $\mathcal{N}=2^*$ (5 次元 $\mathcal{N}=1^*$) 理論と呼ばれるもので，古典的な \SW 曲線自体は以前から知られていました．
この $\widehat{A}_0$ 箙ゲージ理論に対して 2 重量子化の手続きを適用すると，これまでに知られていない新しい W 代数 W$(\widehat{A}_0)$ を得ることが出来ます．
この時 $\mathcal{N}=2^*$ 理論における随伴表現場の質量パラメータが代数の変形パラメータになります．
また，この箙 W 代数の構成をトーラスコンパクト化した 6 次元 $\mathcal{N}=(1,0)$ 理論に適用すると，今度は W 代数の楕円変形を定めることが出来ます．
これは可積分系においてよく知られている有利型・双曲型・楕円型の階層構造がゲージ理論側では 4 次元・ 5 次元・ 6 次元理論に対応していることからの帰結です．

\end{document}